\documentstyle[aps]{revtex}
\begin{document}
\draft
\title{RENORMALIZATION OF GRAVITY AND GRAVITATIONAL WAVES}
\author{Miroslav Pardy}
\address{Department of Theoretical Physics and Astrophysics \\
Masaryk University \\
Kotl\'{a}\v{r}sk\'{a} 2, 611 37 Brno, Czech Republic\\
e-mail:pamir@physics.muni.cz}
\date{\today}
\maketitle
\vspace{50mm}
\begin{abstract}
Strictly respecting the Einstein equations, and supposing space-time is
a medium, we derive the deformation of this medium by gravity.
We derive
the deformation in case of infinite plane, Robertson-Walker manifold,
Schwarzschild manifold and gravitational waves. Some singularities
are removed or changed. We call this procedure renormalization of gravity.
We show that some results following from the classical gravity must be
modified.
\end{abstract}
%\newpage
%\narrowtext

\section{Introduction}

\hspace{3ex}
\baselineskip 15pt

The term renormalization is usually considered as the theoretical
procedure that changes some fundamental constants as charge, mass and
others in such a way that they are in the final form finite
and in agreement with
experiment. While the renormalization in the quantum field theory
can be in a certain sense mastered,
the renormalization procedure in a gravity theory
is in no case
the closed problem. In the Einstein theory of gravity,
one method of the renormalization
consists  for instance in introducing the  cosmological
constant which renormalizes the energy-momentum tensor and
therefore modify the solutions of Einstein equations.

In this article we introduce the special restriction on the
metrical tensor which modifies the consequences of the Einstein equations.
Since the restriction condition has physical meaning, it means that the
consequences of this restriction have also physical meaning.

We know that the main steps in the development of the theoretical view
on the space time was, first, the isolation of space
from time in the Newton period
of physics, and later, the unification of space and time in the special
relativity. The second step was performed
by Lorentz who replaced the Galileian
transformation by the so called Lorentz transformation where
space and time are not isolated and which
was used by Einstein in creation of the special theory
of relativity.

The symbiosis of space and time was expressed elegantly by Minkowski
who introduced the quasieuclidean space-time element where the
space and time are in unity. Minkowski expressed it by
his space-time element $ds$ the square of which is

\begin{equation}
ds^{2} =  c^{2}dt^{2} - dx^{2} - dy^{2}  - dz^{2}.
\label{1}
\end{equation}

Finally, Einstein, in his gravity theory,
replaced the Minkowski space-time element by the
four-dimensional Riemann manifold where the square of the
space-time element was of the form:

\begin{equation}
ds^{2} = g_{\mu\nu}dx^{\mu}dx^{\nu},
\label{2}
\end{equation}
where $g_{\mu\nu}$ is so called the metrical tensor. General relativity
created by Einstein in 1915 is
based on this curved manifold and Einstein equations are
differential equations
for determination of this metrical tensor.

In this article we suppose that the Einstein equations are correct.
These equations are differential equations
for determination of the
metrical tensor $g_{\mu\nu}$ and their form are as follows:

\begin{equation}
R_{\mu\nu} -\frac {1}{2}g_{\mu\nu}R  = \kappa T_{\mu\nu},
\label{3}
\end{equation}
where we here do not consider the additional term
with the cosmological constant which was later
introduced by Einstein in order to regularize some solutions of
his equations. Einstein theory realizes in the mathematical form
the Riemann fundamental idea that space-time
is some form of matter. In other words, there exists in some sense ether. It
means that space is in no case empty and forms some medium.
To deny such medium means ultimately to assume that empty space has no physical
qualities whatever which is in contradiction with the
 general theory of relativity where space is endowed
with physical qualities.
According to the general theory of relativity empty space is unthinkable.
In empty space there not only would be no propagation of
light, but also no possibility of existence for standards of space and time.
If we suppose space-time is a medium, then, it consequently means
that gravitational waves following from the
Einstein theory are quite simply the vibration of space-time
itself \cite{kenyon}.

However, Einstein gives no explanation of the origin of the
metrics, or, metrical tensor.
He "derived" only the nonlinear
equation \cite{chandra} for the metrical tensor and never
explained what is microscopical origin of the metric of space-time.
Einstein supposed that it is adequate that
the metric follows from differential equations as their solution.
However, the metric has an microscopical origin similarly to
situation where the phenomenological thermodynamics
has also the microscopical and statistical origin.

Here we ask a question, what is the
microscopical origin of the metric of space-time.
We postulate that the origin of metric is the
deformation of space-time continuum. We use as an analog
the mechanics of continuum and we apply it to
the space-time medium. The similar approach can be found
in the Tartaglia article, \cite{tartaglia} who also has
considered space-time as a medium, however, without
respecting the Einstein Equations.

In order to explain our ideas let us consider, first,
some continuos
body in some three dimensional Cartesian coordinate
system. The mathematical description of deformation is
as follows \cite{landau1}.

The coordinate of the arbitrary point of the body let be
given by radiusvector ${\bf x}(x^{1}, x^{2}, x^{3})$, where $x^{i}$ are the
Cartesian coordinates. After some deformation the new radiusvector let be
${\bf x}'$. The relative displacement is then ${\bf u}$ where

\begin{equation}
{\bf u} = {\bf x}' - {\bf x}.
\label{4}
\end{equation}

The quantity ${\bf u}$ is so called vector of deformation and obviously

\begin{equation}
dx'^{i}  = dx^{i} + du^{i}.
\label{5}
\end{equation}

While the infinitesimal distance in the nondeformed body is

\begin{equation}
dl^{2} = (dx^{1})^{2}  + (dx^{2})^{2} + (dx^{3})^{2}
=  \delta_{ik}dx^{i}dx^{k},
\label{6}
\end{equation}
for the deformed body we have

\begin{equation}
dl'^{2} = (dx'^{1})^{2}  + (dx'^{2})^{2} + (dx'^{3})^{2}\\
= (dx^{i} + du^{i})^{2} =
(dx^{i})^{2} + 2dx^{i}du^{i} + (du^{i})^{2}.
\label{7}
\end{equation}

Or, with  $du^{i} = \frac {\partial u^{i}}{\partial x^{k}}dx^{k}$

\begin{equation}
dl'^{2} = dl^{2} + u^{ik}dx^{i}dx^{k} = \left(\delta^{ik} + u^{ik}\right)
dx^{i}dx^{k},
\label{8}
\end{equation}
where

\begin{equation}
u^{ik} = \left(\frac {\partial u^{i}}{\partial x^{k}} +
\frac {\partial u^{k}}{\partial x^{i}} + \frac {\partial u^{l}}{\partial
x^{i}} \frac {\partial u^{l}}{\partial x^{k}}\right)
\label{9}
\end{equation}

The last equations are defined for dimension D=3 and Euclidean space.
The analogical equations are evidently valid for Euclidean space with
dimension D = 4. In case of the quasi-euclidean space time with
the dimension D = 4, we can obviously write

\begin{equation}
ds'^{2} = (\eta_{\mu\nu} + u_{\mu\nu})dx^{\mu}dx^{\nu};
\quad \mu, \nu = 0, 1, 2, 3; \quad x^{0} = ct,
\label{10}
\end{equation}
where

\begin{equation}
u_{\mu\nu} = \left(\frac {\partial u_{\mu}}{\partial x_{\nu}} +
\frac {\partial u_{\nu}}{\partial x_{\mu}} +
\frac {\partial u_{\alpha}}{\partial x_{\mu}}
\frac {\partial u^{\alpha}}{\partial x_{\nu}}\right)
= \partial_{\mu}u_{\nu} + \partial_{\nu}u_{\mu} +
\partial_{\mu}u^{\alpha}\partial_{\nu}u_{\alpha}
\label{11}
\end{equation}
and

\begin{equation}
\eta_{\mu\nu} =
\left(\begin{array}{cccc}
1 & 0 & 0 & 0\\
0 & -1 & 0 & 0\\
0 & 0 & -1 & 0\\
0 & 0 & 0 & -1\\
\end{array}\right).
\label{12}
\end{equation}

In such a way, we have for the squared space-time element

\begin{equation}
ds'^{2} = g_{\mu\nu}dx^{\mu}dx^{\nu},
\label{13}
\end{equation}
where

\begin{equation}
g_{\mu\nu} = (\eta_{\mu\nu} + u_{\mu\nu}).
\label{14}
\end{equation}

Equation (14) with definition (11)
excludes gravity with the nonsymmetrial tensor $g_{\mu\nu}$. Or, the
renormalized gravity is symmetrical.

Instead of work with the metrical tensor $g_{\mu\nu}$, we can work with
the tensor of deformation $u_{\mu\nu}$ and we can consider the general theory
of relativity as the four-dimensional theory of some
real deformable medium as a partner of the metrical theory.

\section{The nonrelativistic limit}

The Lagrange function of a point particle with mass $m$ moving in
a potential $\varphi$ is given by the following formula \cite{landau2}:

\begin{equation}
L = -mc^{2} + \frac {mv^{2}}{2} - m\varphi .
\label{15}
\end{equation}

Then, for a corresponding action we have

\begin{equation}
S = \int L dt = -mc \int dt \left(c - \frac {v^{2}}{2c} +
\frac {\varphi}{c} \right) ,
\label{16}
\end{equation}
which ought to be compared with $S = -mc\int ds$. Then,

\begin{equation}
ds = \left(c - \frac {v^{2}}{2c} +  \frac {\varphi}{c} \right) dt.
\label{17}
\end{equation}

With $ d{\bf x} = {\bf v}dt$ and neglecting higher derivative terms, we have

\begin{equation}
ds^{2} = (c^{2} + 2\varphi) dt^{2} - d{\bf x}^{2} =
\left(1 + \frac {2\varphi}{c^{2}}\right)c^{2}dt^{2} - d{\bf x}^{2}.
\label{18}
\end{equation}

The metric determined by this $ds^{2}$ can be be obviously related
to the $u_{\alpha}$ as follows:

\begin{equation}
g_{00} =  1 + 2\partial_{0}u_{0} +
\partial_{0}u^{\alpha}\partial_{0}u_{\alpha} =
1 + \frac {2\varphi}{c^{2}}.
\label{19}
\end{equation}

We can suppose that the time shift caused by the potential is
small and therefore we can neglect the nonlinear term in the last
equation. Then we have

\begin{equation}
g_{00} =  1 + 2\partial_{0}u_{0}  = 1 + \frac {2\varphi}{c^{2}}.
\label{20}
\end{equation}

The elementary consequence of the last equation is

\begin{equation}
\partial_{0}u_{0} =  \frac {\partial u_{0}}{\partial (ct)} =
\frac {\varphi}{c^{2}},
\label{21}
\end{equation}
or,

\begin{equation}
u_{0} = \frac {\varphi}{c}t + const.
\label{22}
\end{equation}

Using $u_{0} = g_{00}u^{0}$, or, $u^{0} = g^{-1}_{00}u_{0} =
\frac {\varphi}{c}$, we get
with

\begin{equation}
u^{0} = ct' - ct,
\label{23}
\end{equation}
the following result (putting the integration constant to zero)

\begin{equation}
t' = t\left(1 + \frac {\varphi}{c^{2}}\right),
\label{24}
\end{equation}
which is the Einstein formula realting time in the zero
gravitational field to time in the gravitational potential
$\varphi$. The shift of light frequency corresponding to the gravitational
potential is, as follows \cite{landau2}.

\begin{equation}
\omega = \omega_{0}\left(1 + \frac {\varphi}{c^{2}}\right).
\label{25}
\end{equation}

So, we have seen that the red shift follows
from our approach immediately,
without application of the Einstein equations.
Now, let us approach the analysis of the physical situations characterized
by the metrics
determined by the Einstein equations. First, let us consider
the metric of the infinite plane.

\section{The infinite plane}

An infinite plane sheet with homogenous mass distribution $\mu$ per unit area
generates the following metrics on it \cite{mizu}:

\begin{equation}
ds^{2} = \left(1 + \frac {8\pi G\mu z}{c^{2}}\right)c^{2}dt^{2}
- dx^{2} -  dy^{2}
- \frac {dz^{2}}{1 + \left(\frac {8\pi G\mu}{c^{2}}\right)z}.
\label{26}
\end{equation}

The corresponding matrix notation of the metric following from
the last equation is:

\begin{equation}
g_{\mu\nu} =
\left(\begin{array}{cccc}
(1+ Az) & 0 & 0 & 0\\
0 & -1 & 0 & 0\\
0 & 0 & -1 & 0\\
0 & 0 & 0 & \left(\frac {-1}{1 + Az}\right)\\
\end{array}\right); \quad A = \frac {8\pi G\mu}{c^{2}}.
\label{27}
\end{equation}

The gravitational potential corresponging to the
metric $g_{\mu\nu}$ is $\varphi = 4\pi G \mu z$
because of equation (19).

It can be observed the singular plane $z = -c^{2}/8\pi G\mu$
in the last metric,which has evidently no physical meaning.
At the same time the metric
is not symmetrical with regard to the $z$-plane. It is not the goal
of this article to solve this specific problem.

It is evident that in case of the infinite plane $u^{0}$ depends only on the
time component and on the $z$ component. In this case of the infinite plane
must be $u^{1} = u^{2} = 0$ and $u^{3}$ is dependent only on the z component.
So, we write:

\begin{equation}
u^{0} = u^{0}(t,z),\quad u^{1} = u^{2} = 0,\quad u^{3} = u^{3}(z).
\label{28}
\end{equation}

By the method analogical to the one in the
preceeding chapter we get the following
equations for the determination of the $u$-components:

\begin{equation}
2\partial_{0}u_{0} + \partial_{0}u^{0}\partial_{0}u_{0} =
\frac {8\pi G\mu z}{c^{2}}
\label{29}
\end{equation}
and

\begin{equation}
-1 + 2\partial_{3}u_{3} + \partial_{3}u_{3}\partial_{3}u_{3} =
- \frac {1}{1 + \frac {8\pi G\mu z}{c^{2}}}.
\label{30}
\end{equation}

Her we can also suppose that the gravitational field generated by the
infinite plane is weak and therefore we can neglect the nonlinear
terms in the last equations. In such a way we have:

\begin{equation}
2\partial_{0}u_{0} = \frac {8\pi G\mu z}{c^{2}} = Az; \quad
 A = \frac {8\pi G\mu }{c^{2}}
\label{31}
\end{equation}
and

\begin{equation}
-1 + 2\partial_{3}u_{3} =  -(1 + Az)^{-1} \approx -1 + Az .
\label{32}
\end{equation}

Using the same procedure as in the preceding chapter, we get in our
linear approximation for the time shift from eq. (29)

\begin{equation}
t' = t\left(1+ Az/2\right) = t\left(1 + \frac {\varphi}{c^{2}}\right),
\label{33}
\end{equation}
where $\varphi$ is the gravitational potential corresponding
to the infinite  plane.
The solution of the equation (32) is
(with $\partial_{3} = \partial/\partial z$)

\begin{equation}
u_{3} = \frac {A}{4}z^{2},
\label{34}
\end{equation}
where we have put the integration constant to zero.

Since

\begin{equation}
u_{3} = g_{33}u^{3} = g_{33}(z'-z),
\label{35}
\end{equation}
we have

\begin{equation}
(z'-z) = g^{-1}_{33}\frac {A}{4}z^{2} \approx  -\frac {Az^{2}}{4},
\label{36}
\end{equation}
or,
\begin{equation}
z' = z\left(1 - \frac {\varphi}{2c^{2}}\right),
\label{37}
\end{equation}
where $\varphi = 4\pi G\mu z$
is the gravitational potential corresponding to the infinite
plane. The integration constant was put to zero with regard to the boundary
condition applied at $z = 0$.

We can see that in case of the gravitating plane there exist not only
the time shift but also the  $z$-coordinate shift. This effect is not
involved in the Will monography \cite{will} on gravitational
experiments and it can be considered as the additional effect
in the general relativity.

\section{The Robertson-Walker metric}

This metric is defined by the following squared space-time element:

\begin{equation}
ds^{2} = c^{2}dt^{2} - R^{2}(t)\left[\frac {dr^{2}}{1-kr^{2}} +
r^{2}d\theta^{2} + r^{2} \sin^{2}\theta d\varphi^{2}\right];
\quad k = -1,0,1.
\label{38}
\end{equation}

The element $ds^{2}$ has the corresponding form in the u-variables:

\begin{equation}
ds^{2} = g_{00}c^{2}dt^{2} - \left[u_{rr}dr^{2} +
u_{\theta \theta}d\theta^{2} +
u_{\varphi\varphi}d\varphi^{2}\right],
\label{39}
\end{equation}
where the nondiagonal elements are equal to zero. According to
Landau \cite{landau1}, $u_{rr} = (\partial/\partial r)u_{r}$, or,

\begin{equation}
u_{rr} = \frac {\partial}{\partial r}u_{r} = \frac {R^{2}(t)}{1-kr^{2}},
\label{40}
\end{equation}
or \cite{dwight},

\begin{equation}
u_{r} = r' - r = R^{2}(t)\frac {1}{2\sqrt k}\ln
\left |\frac {1 + r\sqrt k}{1 - r\sqrt k}\right |.
\label{41}
\end{equation}

The RW metrics with $k = -1$ is called hyperbolic universe, or open universe,
the RW metrics with
$k = 0$ is called flat universe and the RW metrics with $k = 1$
is so called elliptical cosmology, or closed cosmology.
However, we easily see that the last equation for $k = -1$
is not physically meaningful.

At point $k = 0$ there is no singularity. It can be easily seen using
substitution $\varepsilon  = \sqrt{k} r$. Then, for $\varepsilon > 0$ and
$\varepsilon \to 0$, we have from eq. (41):
$J =(1/{2 \sqrt k})\ln [(1 + r\sqrt k)/(1 - r\sqrt k)] =
(r/2\varepsilon)\ln [(1 + \varepsilon)/(1 - \varepsilon)] =$
$(r/2\varepsilon) [2(\varepsilon + \varepsilon^{3}/3 + ...)],
\varepsilon^{2}<1$. And we see that $ J(\varepsilon \to 0)  = r$.

It means that only $k = 0$ and $k = 1$
has physical meaning and so the cosmology is flat or
elliptical. In other words RW universe is flat or closed.
So, in such a way the so called
renormalization of the RW cosmology involves the restriction of
parameters $k$ to $k = 0$ and  $k = 1$,
which is not involved in the original approach to cosmology.

\section{The Schwarzschild metric}

The squared space-time element of the Schwarzschild space-time is of the
form:

\begin{equation}
ds^{2} = \left(1 - \frac {r_{g}}{r}\right)c^{2}dt^{2} - \frac {dr^{2}}{1 -
\frac{r_{g}}{r}} - r^{2}(\sin^{2}\Theta d\varphi^{2} + d\Theta^{2}),\quad
r_{g} = \frac {2MG}{c^{2}}.
\label{42}
\end{equation}

It means that for the u-components we have:

\begin{equation}
g_{00} =   1 + 2\partial_{0}u_{0} = \left(1 -  \frac {r_{g}}{r}\right),
\quad u_{rr} =  \frac {\partial}{\partial r} u_{r} =
\frac {1}{1 - \frac{r_{g}}{r}}.
\label{43}
\end{equation}

After elementary integration we get for the t-displacement

\begin{equation}
u^{0} =  c(t' - t) =  - \frac {r_{g}}{2r}ct + const,
\label{44}
\end{equation}
or, (we put the constant to zero)

\begin{equation}
t'= \left(1  - \frac {r_{g}}{2r}t\right) =
\left(1 + \frac {\varphi}{c^{2}}\right); \quad \varphi = -\frac {MG}{r}.
\label{45}
\end{equation}
corresponds exactly to the formula of the time
dilation with the Newton potential $\varphi = -\frac {GM}{r}$.

For r-displacement it is

\begin{equation}
u_{r} = r' - r = \int \frac {r dr}{r-r_{g}} = r + r_{g}\ln|r-r_{g}| + const.
\label{46}
\end{equation}

We observe that the left and right limit  of $u_{r}$ for $r \to r_{g}$are
identical in contradistinction with the limit in the original
Schwarzschild metric. In oder words,

\begin{equation}
\lim_{r \to r_{g}+} \ln|r-r_{g}| = \lim_{r \to r_{g}-} \ln|r-r_{g}|.
\label{47}
\end{equation}

While the singularity in the original metric is hyperbolical, the singularity
after the $u$-mechanism is only logarithmical. So, the renormalization
in this case is soft.

\section{The weak gravitational waves}

The weak gravitational field is defined by the  metric

\begin{equation}
g_{\mu\nu} = \eta_{\mu\nu} + h_{\mu\nu},
\label{48}
\end{equation}
where

\begin{equation}
|h_{\mu\nu}| \ll 1.
\label{49}
\end{equation}

It is possible to show that in this case the equation for these
weak gravitational waves is the linear wave equation.

\begin{equation}
\Box^{2}h_{\mu\nu} = 0
\label{50}
\end{equation}
with additional constrain following from the conservation of laws,

\begin{equation}
\partial_{\mu}h^{\mu}_{\nu} = \frac {1}{2}\partial_{\nu}h^{\sigma}_{\sigma}.
\label{51}
\end{equation}

The gravitational waves in empty space are
determined  as a solution of
the wave equation and has the well known form:

\begin{equation}
h_{\mu\nu} = a_{\mu\nu}\cos({\bf kx} - \omega t),
\label{52}
\end{equation}
where $a_{\mu\nu}$ is the tensor amplitude and it means
that the differential equations for the $u_{\mu}$-functions must be
related to $h_{\mu\nu}$ in the following form:

\begin{equation}
a_{\mu\nu}\cos({\bf kx} - \omega t)  =
\partial_{\mu}u_{\nu} + \partial_{\nu}u_{\mu},
\label{53}
\end{equation}
where we have neglected the the nonlinear $u$-term  on the
right side of the last equation, because it is of the second order
and it cannot be in principle related in this approximation
to the terms of the first order.

The solution of the last equation forms simple mathematical
problem which can be solved easily supposing

\begin{equation}
u_{\mu} = u_{\mu}(x,y,z,t)
\label{54}
\end{equation}

Let us look for the wave propagating in direction $x^{3} = z$  with
${\bf k} \equiv (0,0,k)$.
In this case, it is well known that only the nonzero term $h_{\alpha\beta}$
are the following ones:

\begin{equation}
a_{11} = - a_{22}; \quad a_{12} = a_{21}
\label{55}
\end{equation}

Then, using relations (55) in eq. (\ref{53}) we get

\begin{equation}
A\cos( kz - \omega t)  = 2\frac {\partial u_{1}}{\partial x};
\quad A = a_{11}
\label{56}
\end{equation}

\begin{equation}
-A\cos(kz - \omega t)  = 2\frac {\partial u_{2}}{\partial y};
\quad A = a_{11}
\label{57}
\end{equation}

\begin{equation}
B\cos(kz - \omega t)  = \frac {\partial u_{1}}{\partial y} +
\frac {\partial u_{2}}{\partial x};
\quad B = a_{12}.
\label{58}
\end{equation}

The solution of the last equations can be obtained by the
elementary integration in the following form
(with $\varphi = (kz - \omega t)$):

\begin{equation}
u_{1} = \frac {A}{2}x\cos\varphi  + \alpha(y, z, t);
\quad u_{2} = \frac {-A}{2}y\cos\varphi + \beta(x, z, t),
\label{59}
\end{equation}
where $\alpha$ and $\beta$  are functions which can be determined from
inserting eq. (59) to the eq. (58). Or,

\begin{equation}
\frac {\partial \alpha}{\partial y} + \frac {\partial \beta}{\partial x} =
B\cos \varphi ,
\label{60}
\end{equation}
and the solution of the last equation is of the form:

\begin{equation}
\alpha = \frac {B}{2}y\cos\varphi;\quad
\beta = \frac {B}{2}x\cos\varphi
\label{61}
\end{equation}

Now, we can write the four vector corresponding to the weak gravitational
wave spreading in the direction of the z-axis in the form:

\begin{equation}
u =
\left(\begin{array}{c}
u_{0} \\
u_{1} \\
u_{2}\\
u_{3}
\end{array}\right)
=
\frac {1}{2}
\left(\begin{array}{c}
0 \\
Ax + By \\
-Ay + Bx\\
0
\end{array}\right)\cos\varphi; \quad \varphi = (kz - \omega t)
\label{62}
\end{equation}

The constants  $A$ and $B$ must be chosen in such a way that $A = B$
because of the equivalence of the of $x$-coordinates with
$y$-coordinates in free space.

\section{Discussion}

We have defined general relativity and gravitation as a
deformation of a medium called space-time. We have used specific
equation which relates Einstein metric to the displacement
of points of the medium and applied it to the some gravitating
systems, such as infinite plane desk, Schwartzschild universe,
Robertson walker universe, and gravitational waves. It is evident that our
method can be also applied to the further systems,
such as the rotating plane desk, anti de Sitter space-time,
the Reissner-Nordstrom space-time, the Newmann et al. space-time,
the Kerr space-time and
so on. Such aplication can be considered as interesting problems
and will have certainly meaning in the future gravitational physics.

The future measurement of the gravitational
waves, performed by the LIGO project \cite{ligo},
VIRGO project \cite{virgo},
GEO 600 project \cite{geo} and TAMA 300 project \cite{tama}
will evidently have also the positive impact on the development of
gravitational physics.

\end{document}